\documentstyle[12pt,epsfig,amssymb]{article}
\begin{document}
\begin{flushright}
CERN-TH/97-284 \\
\end{flushright}

\begin{center}
{\LARGE\bf Parton Distributions from Photoproduction in Polarized 
$ep$ Scattering at HERA}
\vspace{0.6cm}

\renewcommand{\thefootnote}{\fnsymbol{footnote}}
\setcounter{footnote}{0}
{\Large Werner Vogelsang\footnote{Invited talk presented at the workshop 
`Deep Inelastic Scattering off Polarized Targets:  Theory Meets Experiment', 
Zeuthen, Germany, September 1-5, 1997}}
\renewcommand{\thefootnote}{\fnsymbol{roman}}
\setcounter{footnote}{0}

\vspace*{0.4cm}
{\it Theory Division, CERN, CH-1211 Geneva, 
Switzerland}\\

\vspace*{0.3cm}

\end{center}

\begin{abstract}
We study photoproduction of jets and single-inclusive hadrons 
in a polarized $ep$ collider mode of HERA at $\sqrt{s}\approx 300$ GeV,
examining the sensitivity of the cross sections and their asymmetries 
to the proton's polarized gluon distribution and to the completely unknown 
parton distributions of polarized photons. We also present for the 
first time the NLO corrections to the direct part of polarized
single-inclusive hadron photoproduction.
\end{abstract}
\vspace*{-0.4cm}

\section{Introduction}
\noindent
Among the various conceivable options for future HERA upgrades is the
idea to longitudinally polarize its proton beam~\cite{sch} which, 
when combined with the already operative longitudinally polarized electron 
(positron) beam, results in a polarized version of the usual HERA collider 
with $\sqrt{s}\approx 300$ GeV. A typical conservative value for the 
integrated luminosity in this case should be 100 $\mbox{pb}^{-1}$.

HERA has already been very successful in pinning down 
the proton's unpolarized gluon distribution $g(x,Q^2)$. Several processes 
have been studied which have contributions from $g(x,Q^2)$ already in the 
lowest order, such as (di)jet, inclusive hadron, and heavy flavour production. 
Since events at HERA are concentrated in the region $Q^2 \rightarrow 0$, the 
processes have first and most accurately been studied in 
photoproduction~[2-7]. As is well-known, in this case the 
(quasi-real) photon will not only interact in a direct (`point-like') way, 
but can also be resolved into its hadronic structure. HERA photoproduction 
experiments like~[2-7] have not merely established evidence for the existence 
of such a resolved contribution, but have also been precise enough to 
improve our knowledge about the parton distributions, $f^{\gamma}$, of the 
photon.

Given the success of such unpolarized photoproduction experiments at HERA, 
it seems most promising~\cite{wir} to closely examine the same processes for 
the situation with longitudinally polarized beams with regard to their 
sensitivity to the proton's polarized gluon distribution $\Delta g$, which is
still one of the most interesting, but least known, quantities in
`spin-physics'. Recent next-to-leading (NLO) studies of polarized
DIS~[9-12] show that the $x$-shape of $\Delta g$ seems to be 
hardly constrained at all by the present DIS data, even though
a tendency towards a sizeable positive {\em total} gluon polarization,
$\int_0^1 \Delta g(x,Q^2=4 \; \mbox{GeV}^2) dx \gtrsim 1$, was found
\cite{grsv,bfr,gs}. Furthermore, polarized photoproduction experiments 
may in principle allow to not only determine the parton, in
particular gluon, content of the polarized {\em proton}, but also 
that of the longitudinally polarized {\em photon} which is completely
unknown so far. Since a measurement of, e.g., the photon's spin-dependent
structure function $g_1^{\gamma}$ in polarized $e^+ e^-$ collisions is not
planned in the near future, 
\nopagebreak[4]
polarized HERA could play a unique role here, 
even if it should only succeed in establishing the very {\em existence} 
of a resolved contribution to polarized photon-proton reactions.
\pagebreak[4]

Our contribution is organized as follows: In the next section we collect 
the necessary ingredients for our calculations. In sec.~3 we will discuss 
at leading order (LO) the most promising photoproduction reactions, 
namely (di)jet and single-inclusive hadron production. Part of this 
section is taken from~\cite{wir}. Sec.~4 will then present for the first 
time the NLO corrections to the direct part of the latter process. 
More details on the results presented in sec.~4 will be published in~\cite{dv}.
\section{Spin-dependent Parton Distributions of the Proton and the Photon}
\noindent
Our main calculations will be performed at LO, as the
NLO corrections to the spin-dependent parts of the photoproduction
processes we are interested in are usually not yet available.  
This implies use of LO parton distributions, which have been provided 
in the analyses~\cite{grsv,gs} of recent polarized DIS data. Both papers 
give various LO sets which mainly differ in the $x$-shape of the 
polarized gluon distribution. We will choose the LO `valence' set of the 
`radiative parton model analysis'~\cite{grsv}, which 
corresponds to the best-fit result of that paper, along with two other sets 
of~\cite{grsv} which are based on either assuming $\Delta g (x,\mu^2) = 
g(x,\mu^2)$ or $\Delta g(x,\mu^2)=0$ at the low input scale $\mu$ of 
\cite{grsv}, where $g(x,\mu^2)$ is the unpolarized LO GRV~\cite{grv} input 
gluon distribution. These two sets will be called `$\Delta g=g$ input' and 
`$\Delta g=0$ input' scenarios, respectively. The gluon of set C of~\cite{gs}
is qualitatively different since it has a substantial negative polarization 
at large $x$. We will therefore also use this set in our calculations.
For illustration, we show in Fig.~1 the gluon distributions  
of the four different sets of parton distributions we will use, taking a 
typical scale $Q^2=10$ GeV$^2$. Keeping in mind that all four LO sets provide 
very good descriptions of the present polarized DIS data, 
it becomes obvious that the data indeed do not seem to be able to 
significantly constrain the $x$-shape of $\Delta g(x,Q^2)$.

In the case of photoproduction the electron just
serves as a source of quasi-real photons which are radiated according 
to the Weizs\"{a}cker-Williams spectrum. The photons can then 
interact either directly or via their partonic structure (`resolved' 
contribution). In the case of longitudinally polarized electrons, the 
resulting photon will be longitudinally (more precisely, circularly)
polarized and, in the resolved case, the {\em polarized} (helicity-weighted) 
parton distributions of the photon, $\Delta f^{\gamma}(x,Q^2)$, enter the 
calculations. Thus one can define the effective polarized parton densities 
at the scale $M$ 
in the longitudinally polarized electron via\footnote{We 
include here the additional definition 
$\Delta f^{\gamma} (x_{\gamma},M^2) \equiv \delta (1-x_{\gamma})$ for the 
direct (`unresolved') case.}
\begin{equation}  \label{elec}
\Delta f^e (x_e,M^2) = \int_{x_e}^1 \frac{dy}{y} \Delta P_{\gamma/e} (y)
\Delta f^{\gamma} (x_{\gamma}=\frac{x_e}{y},M^2) \; 
\end{equation}
($f=q,g$) where $\Delta P_{\gamma/e}$ is the polarized Weizs\"{a}cker-Williams
spectrum for which we will use 
\begin{equation}  \label{weiz}
\Delta P_{\gamma/e} (y) = \frac{\alpha_{em}}{2\pi} \left[ 
\frac{1-(1-y)^2}{y} \right] \ln \frac{Q^2_{max} (1-y)}{m_e^2 y^2} \; ,
\end{equation}
with the electron mass $m_e$. For the time being, it seems most 
sensible to follow as closely as possible the analyses successfully 
performed in the unpolarized case, which implies to introduce 
\begin{figure}[t]
\begin{center}
\vspace*{-1cm}
\hspace*{0.5cm}
\epsfig{file=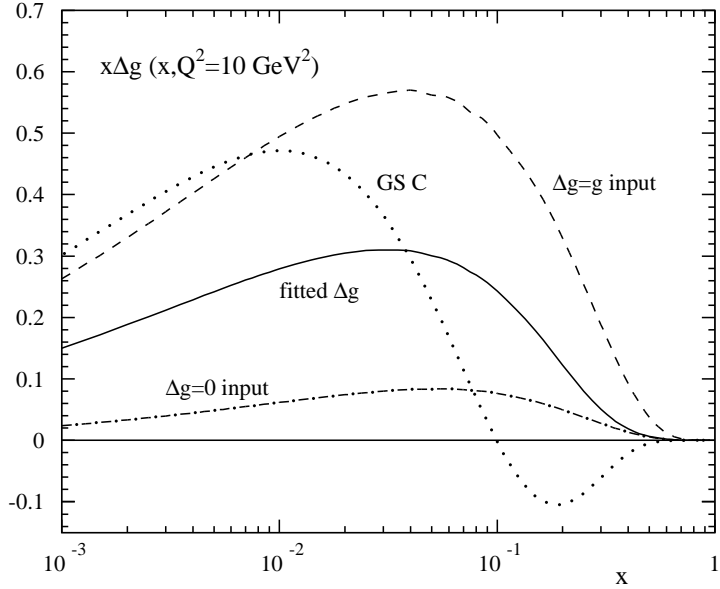,width=11cm}
\vspace*{-0.9cm}
\caption{\sf Gluon distributions at $Q^2=10\,{\mbox{GeV}}^{\:2}$ of the four 
LO sets of polarized parton distributions used in this paper. The dotted line 
refers to set C of~\cite{gs}, whereas the other distributions are taken 
from~\cite{grsv} as described in the text.}
\vspace*{-0.7cm}
\end{center}
\end{figure}
the same 
kinematical cuts. As in~\cite{jet1ph,jet2ph,kramer} we will use
an upper cut\footnote{In H1 analyses of HERA photoproduction
data~\cite{jet1h1,jet2h1} the cut $Q^2_{max}=0.01$ GeV$^2$ is used 
along with slightly different $y$-cuts as compared to the corresponding 
ZEUS measurements~\cite{jet1ph,jet2ph}, which leads to smaller rates.} 
$Q^2_{max}=4$ GeV$^2$, and the $y$-cuts $0.2 \leq y \leq 0.85$ 
(for single-jet~\cite{jet1ph} production) and $0.2 \leq y \leq 0.8$ 
(for dijet~\cite{jet2ph} and single-inclusive hadron \cite{ihh1,ihzeus} 
production) will be imposed. 

The polarized photon structure functions $\Delta f^{\gamma} (x_{\gamma},M^2)$ 
in (\ref{elec}) are completely unmeasured so far, so that 
models for them have to be invoked. To obtain a realistic estimate for the 
theoretical uncertainties in the polarized photonic parton densities
two very different scenarios were considered in~\cite{gvg} assuming 
`maximal' ($\Delta f^{\gamma}(x,\mu^2)=f^{\gamma}(x,\mu^2)$) or `minimal' 
($\Delta f^{\gamma}(x,\mu^2)=0$) saturation of the fundamental positivity 
constraints $|\Delta f^{\gamma}(x,\mu^2)| \leq f^{\gamma}(x,\mu^2)$ at the
input scale $\mu$ for the QCD evolution. Here $\mu$ and the unpolarized 
photon structure functions $f^{\gamma}(x,\mu^2)$ were adopted from the 
phenomenologically successful radiative parton model predictions in 
\cite{grvg}. The results of these two extreme approaches are presented in 
Fig.~2 in terms of the photonic parton asymmetries $A_f^{\gamma} \equiv 
\Delta f^{\gamma}/f^{\gamma}$, evolved to $Q^2=30$ GeV$^2$ in LO. An ideal 
aim of measurements in a polarized collider mode of HERA would of course be 
to determine the $\Delta f^{\gamma}$ and to see which ansatz is more 
realistic. The sets presented in Fig.~2, which we 
will use in what follows, should in any case be sufficient to study 
the sensitivity of the various cross sections to the $\Delta f^{\gamma}$,
but also to see in how far they influence a determination of $\Delta g$.
We note that in \cite{svnlo} we have extended our studies of the polarized 
photon structure also to NLO. 
\begin{figure}[ht]
\begin{center}
\vspace*{-1cm}
\hspace*{0.4cm}
\epsfig{file=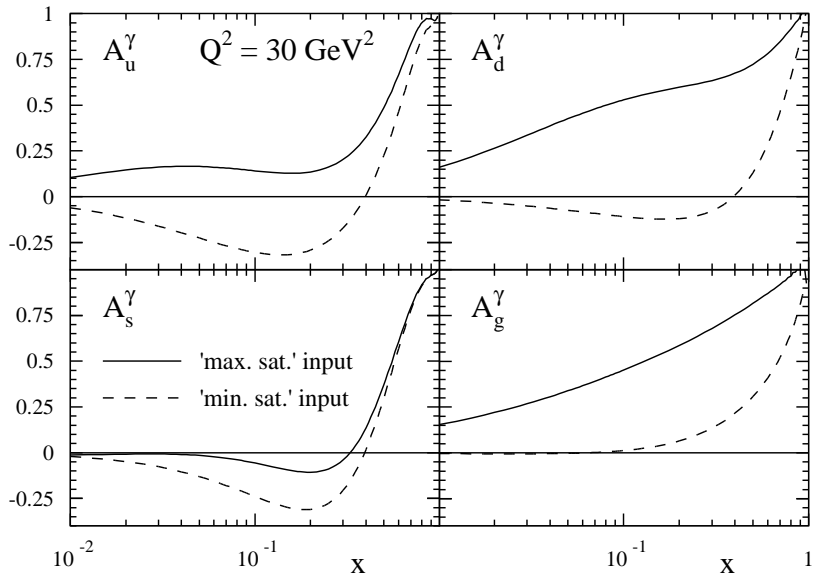,width=12cm}
\vspace*{-0.55cm}
\caption{\sf Photonic LO parton asymmetries $A_f^{\gamma}\equiv 
\Delta f^{\gamma}/f^{\gamma}$ at $Q^2=30\,{\mbox{GeV}}^{\,2}$ 
for the two scenarios considered in~\cite{gvg} (see text). The 
unpolarized LO photonic parton distributions were taken from~\cite{grvg}.}
\vspace*{-0.7cm}
\end{center}
\end{figure}

We finally note that in what follows a polarized cross section will always
be defined as
\begin{equation} 
\Delta \sigma \equiv \frac{1}{2} \Big[ \sigma (++)-\sigma (+-) \Big] \; ,
\end{equation}
the signs denoting the helicities of the scattering particles. The
corresponding unpolarized cross section is given by taking the sum 
instead, and the cross section asymmetry is $A\equiv \Delta \sigma/\sigma$.
Whenever calculating an asymmetry $A$, we will 
use the LO GRV parton distributions for the proton~\cite{grv} and the 
photon~\cite{grvg} to calculate the unpolarized cross section.
For consistency, we will employ the LO expression for the strong coupling 
$\alpha_s$ with~\cite{grsv,gs,gvg} $\Lambda_{QCD}^{(f=4)}=200$ MeV for 
four active flavours.
\section{Photoproduction Reactions at Polarized HERA}
\noindent
The generic LO cross section formula for the photoproduction of a single jet 
with transverse momentum $p_T$ and cms-rapidity $\eta$ in polarized $ep$ 
collisions reads:
\begin{equation} \label{wqc}
\frac{d^2 \Delta \sigma}{dp_T d\eta} =  \sum_{f^e,f^p,c} 
\Delta f^e (x_e,M^2) \otimes \Delta f^p (x_p,M^2) \otimes 
\frac{d^2 \Delta \hat{\sigma}^{f_e f_p \rightarrow cd}}{dp_T d\eta} \; ,
\end{equation}
where $\otimes$ denotes a convolution and the sum is running over all 
properly symmetrized $2\rightarrow 2$ subprocesses for the direct 
($\gamma b\rightarrow cd$, $\Delta f^e (x_e,M^2) \equiv 
\Delta P_{\gamma/e}(x_e)$) and resolved ($ab\rightarrow cd$) cases. When only 
light flavours are involved, the corresponding differential helicity-dependent 
LO subprocess cross sections can be found in~\cite{bab}. In all following 
predictions we will deal with the charm contribution to the cross section 
by including charm only as a {\em{final}} state particle produced via the 
subprocesses $\gamma g \rightarrow c\bar{c}$ (for the direct part) and $gg 
\rightarrow c\bar{c}$, $q\bar{q} \rightarrow c\bar{c}$ (for the resolved 
part). For the values of $p_T$ considered it turns out that the finite 
charm mass 
can be safely neglected in these subprocess cross sections. In (\ref{wqc}), 
$\hat{s} \equiv x_e x_p s$ and $M$ is the factorization/renormalization scale 
for which we will use\footnote{The scale dependence of the theoretical LO
predictions for the spin asymmetries -- which are the quantities relevant in 
experiments -- turns out to be rather weak, for a discussion see~\cite{wir}.}
$\; M=p_T$. The $\Delta f^p$ stand for the 
polarized parton distributions of the proton. Needless to say that we  
obtain the corresponding unpolarized LO jet cross section $d^2 \sigma/
dp_T d\eta$ by using LO unpolarized parton distributions and subprocess cross 
sections in (\ref{wqc}). 

It appears very promising~\cite{wir} to study the $\eta_{LAB}$-distribution of 
the cross section and the asymmetry, where $\eta_{LAB}$ is the laboratory
frame rapidity, related to $\eta$ via $\eta \equiv \eta_{cms} = \eta_{LAB} -
\frac{1}{2} \ln (E_p/E_e)$. As usual, $\eta_{LAB}$ is defined to be positive
in the proton forward direction. The crucial point is that for negative 
$\eta_{LAB}$ the main contributions are expected to come from the region 
$x_{\gamma} \rightarrow 1$ and thus mostly from the direct piece at 
$x_{\gamma}=1$. To investigate this, Fig.~3 shows our results for the 
single-inclusive jet cross section and its asymmetry vs. $\eta_{LAB}$ 
and integrated over $p_T>8$ GeV for the four sets of the polarized 
proton's parton distributions. For Figs.~3a,b we have used the `maximally' 
saturated set of polarized photonic parton densities, whereas Figs.~3c,d 
correspond to the `minimally' saturated one. Comparison of Figs.~3a,c or
3b,d shows that indeed the direct contribution clearly dominates for 
$\eta_{LAB} \leq -0.5$, where also differences between the polarized gluon 
distributions of the proton show up clearly. Furthermore, the cross sections 
are generally large in this region with asymmetries of a few percents. At 
positive $\eta_{LAB}$, we find that the cross section is 
dominated by the resolved contribution and is therefore sensitive to
the parton content of both the polarized proton {\em and} the photon.
This means that one can only learn something about the polarized photon 
structure functions if the polarized parton distributions 
\begin{figure}[t]
\begin{center}
\vspace*{-0.9cm}
\hspace*{0.5cm}
\epsfig{file=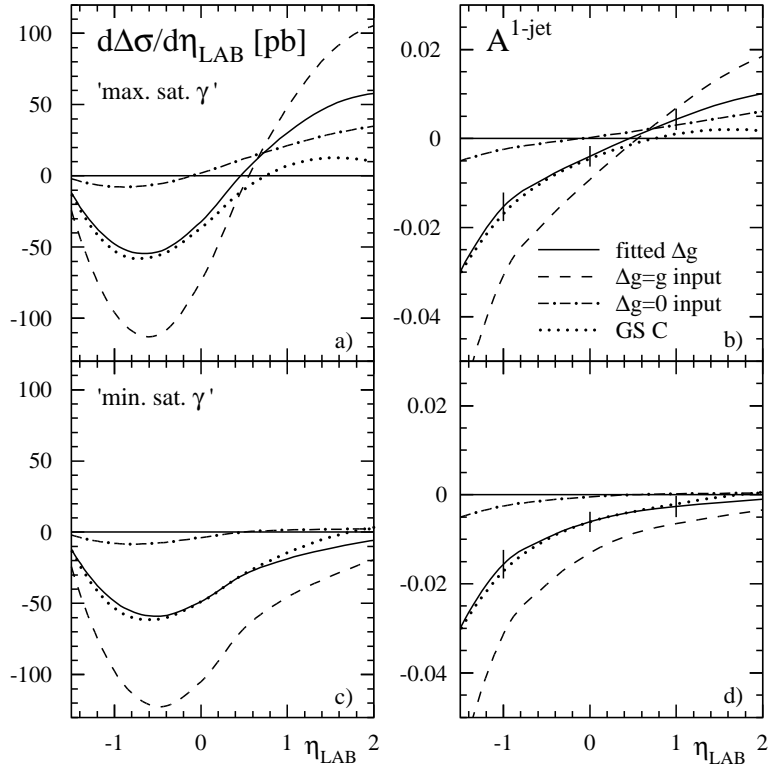,width=11.5cm}
\vspace*{-0.7cm}
\caption{\sf {\bf a:} $\eta_{LAB}$-dependence of the polarized single-jet 
inclusive photoproduction cross section in $ep$-collisions at HERA, integrated
over $p_T > 8$ GeV. The resolved contribution to the cross section has 
been calculated with the `maximally' saturated set of polarized photonic 
parton distributions. {\bf b:} Asymmetry corresponding to {\bf a}. 
{\bf c,d:} Same as {\bf a,b}, but for the `minimally' saturated set of 
polarized photonic parton distributions.}
\vspace*{-0.7cm}
\end{center}
\end{figure}
of the proton are 
already known to some accuracy or if an experimental distinction between 
resolved and direct contributions can be achieved. We note that the dominant 
contributions to the resolved part at large $\eta_{LAB}$ are driven by the 
polarized photonic {\em gluon} distribution $\Delta g^{\gamma}$. 
We have included in the asymmetry plots in Figs.~3b,d the expected 
statistical errors $\delta A$ at HERA which can be estimated from 
\begin{equation}  \label{aerr}
\delta A = \frac{1}{P_e P_p \sqrt{{\cal L} \sigma \epsilon}} \; ,
\end{equation}
where $P_e$, $P_p$ are the beam polarizations, ${\cal L}$ is the integrated 
luminosity and $\epsilon$ the jet detection efficiency, for which we assume
$P_e * P_p=0.5$, ${\cal L}=100$/pb and $\epsilon=1$. From the results it 
appears that a measurement of the proton's $\Delta g$ should be possible from 
single-jet events at negative rapidities where the contamination from the 
resolved contribution is minimal. 

In the unpolarized case, an experimental criterion for a distinction  
between direct and resolved contributions has been introduced~\cite{jeff} and 
used~\cite{jet2ph} in the case of dijet photoproduction at HERA. We will now 
adopt this criterion for the polarized case to see whether it would enable a
further access to $\Delta g$ and/or the polarized photon structure functions. 
The generic expression for the polarized cross section $d^3 \Delta \sigma/dp_T 
d\eta_1 d\eta_2$ for the photoproduction of two jets with laboratory system 
rapidities $\eta_1$, $\eta_2$ has a form analogous to (\ref{wqc}).
Here one has 
\begin{equation}
x_e \equiv \frac{p_T}{2 E_e} \left( e^{-\eta_1} + e^{-\eta_2} \right)\;\; , \;
x_p \equiv \frac{p_T}{2 E_p} \left( e^{\eta_1} + e^{\eta_2} \right) \; ,
\end{equation}
where $p_T$ is the transverse momentum of one of the two jets (which balance
each other in LO).
Following~\cite{jet2ph}, we will integrate over the cross section to obtain 
$d\Delta \sigma/d\bar{\eta}$, where $\bar{\eta} \equiv (\eta_1 + \eta_2)/2$.
Furthermore, we will apply the cuts~\cite{jet2ph}
$|\Delta \eta| \equiv |\eta_1-\eta_2| \leq 0.5 \; , \;\; 
p_T>6 \; \mbox{GeV}$.
The important point is that measurement of the jet rapidities allows 
for fully reconstructing the kinematics of the underlying hard subprocess
and thus for determining the variable~\cite{jet2ph}
\begin{equation}
x_{\gamma}^{OBS} = \frac{\sum_{jets} p_T^{jet} e^{-\eta^{jet}}}{2yE_e} \; ,
\end{equation} 
which in LO equals $x_{\gamma}=x_e/y$ with $y$ as before being the 
fraction of the electron's energy taken by the photon. Thus it becomes
possible to experimentally select events at large $x_{\gamma}$, 
$x_{\gamma} > 0.75$~\cite{jeff,jet2ph}, 
hereby isolating the {\em direct} contribution to 
the cross section with just a rather small contamination from resolved 
processes. Conversely, the events with $x_{\gamma}\leq 0.75$ will represent 
the resolved part of the cross section. This procedure should therefore be 
ideal to extract $\Delta g$ on the one hand, and examine the polarized 
photon structure functions on the other.

Fig.~4 shows the results~\cite{wir} for the direct part of the cross section 
according to the above selection criteria. The contributions from the 
resolved subprocesses have been included, using the `maximally' 
saturated set of polarized photonic parton densities. They turn out
to be non-negligible but, as expected, subdominant. More importantly,
due to the constraint $x_{\gamma}>0.75$ they are determined by the 
polarized quark, in particular the $u$-quark, distributions in the photon, 
which at large $x_{\gamma}$ are equal to their unpolarized counterparts as 
a result of the $Q^2$-evolution (see Fig.~2), rather {\em independently} of 
the hadronic input chosen. Thus the uncertainty coming from the polarized 
photon structure is minimal here and under control.
\begin{figure}[p]
\begin{center}
\vspace*{-1.6cm}
\hspace*{0.6cm}
\epsfig{file=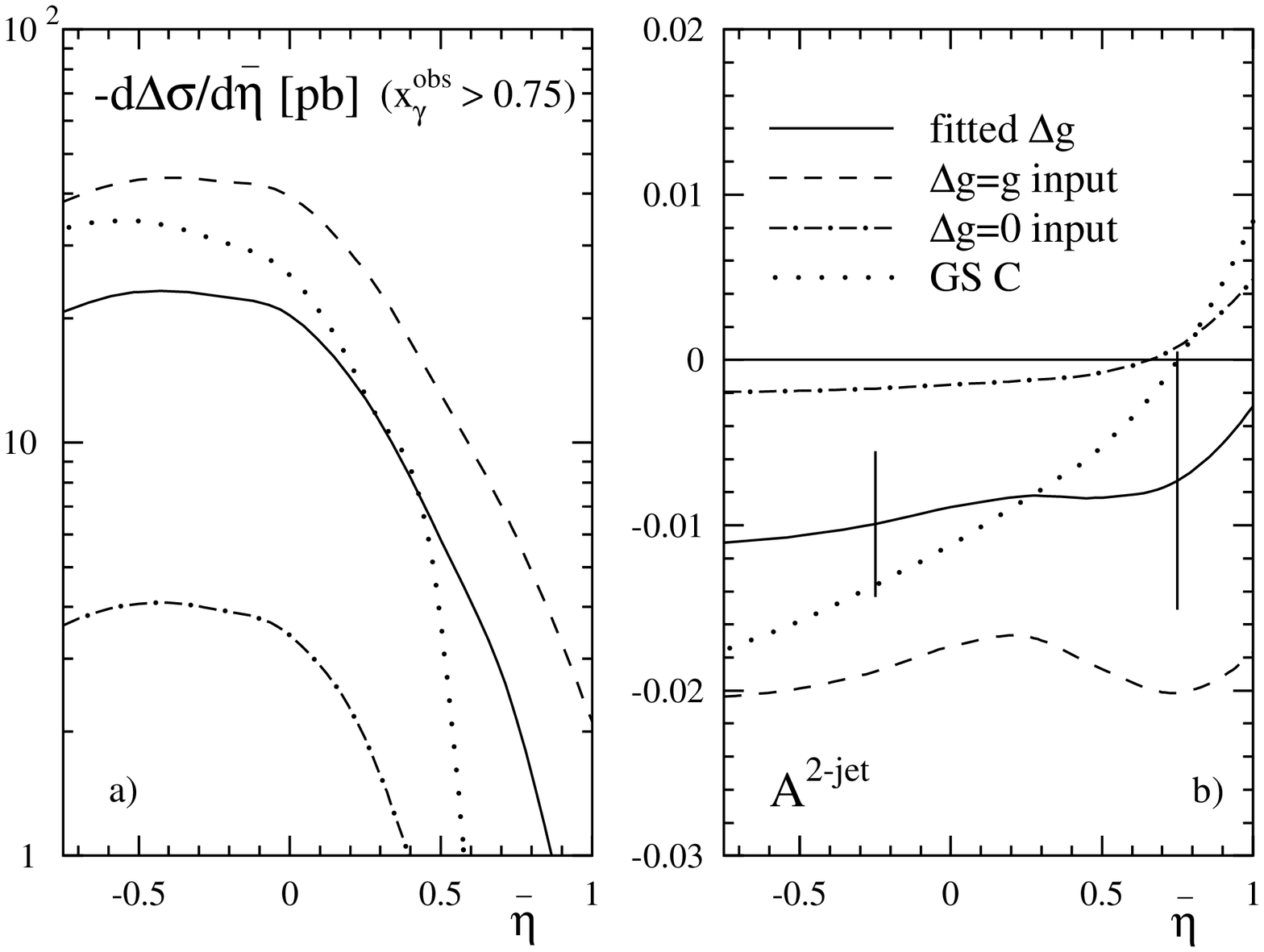,width=11cm}
\vspace*{-0.4cm}
\caption{\sf {\bf a:} $\bar{\eta}$-dependence of the `direct' part 
($x_{\gamma}^{OBS}>0.75$) of the polarized two-jet photoproduction cross 
section in $ep$-collisions at HERA for the four different sets of polarized 
parton distributions of the proton. {\bf b:} Asymmetry corresponding to 
{\bf a}. The expected statistical errors indicated by the bars have been 
calculated according to (\ref{aerr}) and as explained in the text.}
%
\hspace*{1.3cm}
\epsfig{file=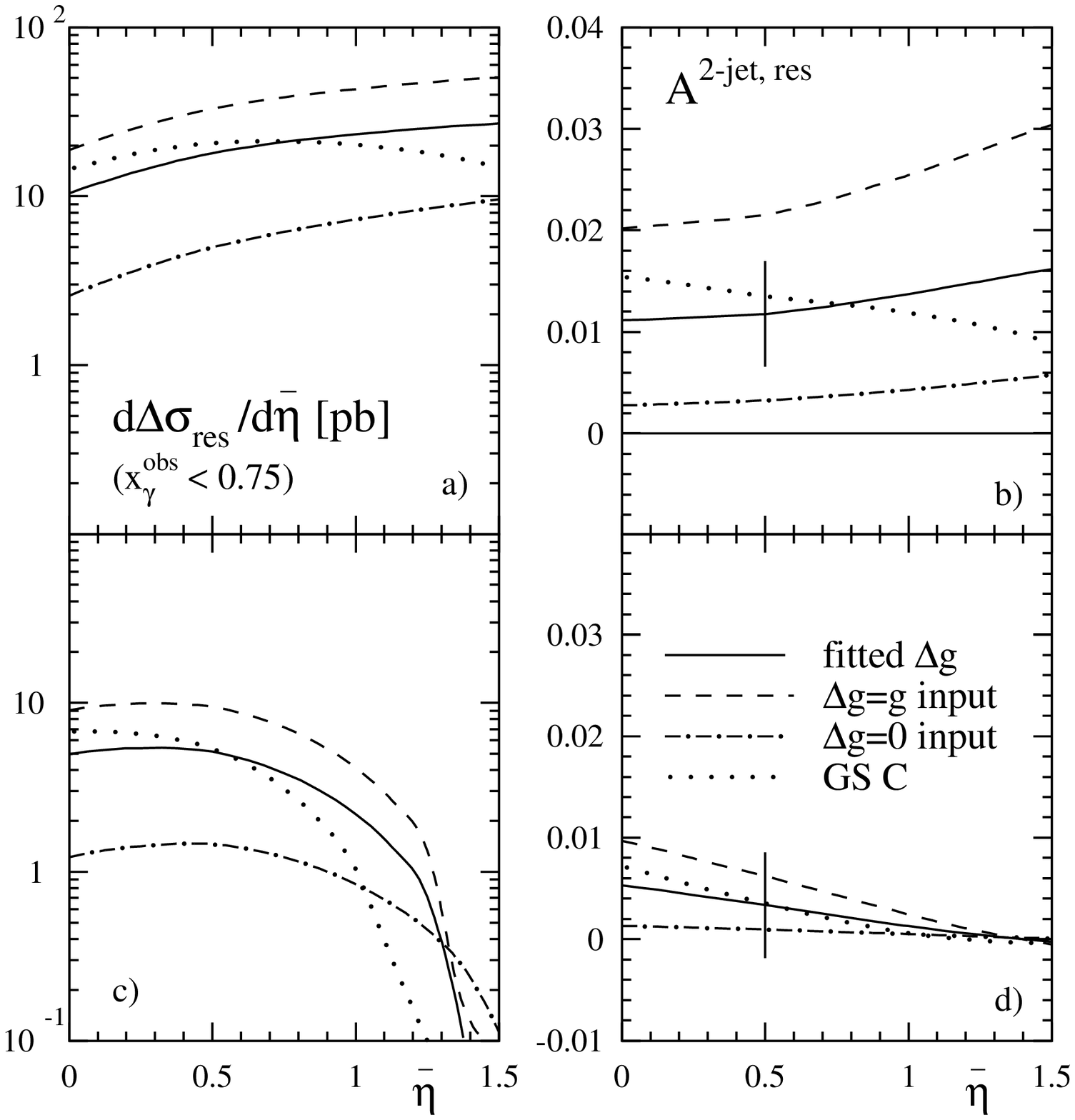,width=11cm}
\vspace*{-0.4cm}
\hspace*{0.5cm}
\caption{\sf Same as Fig.~4, but for the resolved part of the cross section, 
defined by $x_{\gamma}^{OBS}\leq 0.75$ (see text). For {\bf a,b} the 
`maximally' saturated set of polarized photonic parton distributions has 
been used and for {\bf c,d} the `minimally' saturated one.}
\end{center}
\end{figure}
As becomes obvious from Fig.~4, the cross sections are fairly large over the 
whole range of $\bar{\eta}$ displayed and very sensitive to the shape 
{\em and} the size of $\Delta g$ with, unfortunately, not too sizeable 
asymmetries as compared to the statistical errors for ${\cal L}=100$/pb. 
A measurement of $\Delta g$ thus appears to be possible under the imposed 
conditions only if luminosities clearly exceeding $100$/pb can be reached. 
Fig.~5 displays the same results, but now for the resolved 
contribution with $x_{\gamma} \leq 0.75$ for the `maximally' saturated set 
(Figs.~5a,b) and the `minimally' saturated one (Figs.~5c,d). As expected, 
the results depend on the parton content of both the polarized photon and 
the proton, which implies that again the latter has to be known to some 
accuracy to allow for the extraction of some 
information on the polarized photon structure. 
We emphasize that the experimental finding of a non-vanishing 
asymmetry here would establish at least the definite existence of a resolved 
contribution to the polarized cross section. 

At a first glance, single-inclusive production of charged hadrons
appears less interesting than jet production, as the 
cross section for producing a definite hadron at a given $p_T$ will always be 
smaller than the one for a jet. On the other hand, in case of inclusive 
hadrons one can obviously go experimentally to $p_T$ much smaller than the
$p_T^{min}=8$ GeV employed in our jet studies. Moreover, in the unpolarized
case single-inclusive hadron production was successfully studied 
experimentally at HERA prior to jets~\cite{ihh1,ihzeus}. The expression for 
the cross section for single-inclusive hadron production is similar to the 
one in (\ref{wqc}), but comprises an additional convolution with the 
function $D_c^h$ describing the fragmentation of particle $c$ into the
hadron $h$. For the $D_c^h$ we will use the LO fragmentation functions
of~\cite{bkk} which yield a good description of the unpolarized HERA inclusive
hadron data~\cite{ihh1,ihzeus}. Figs.~6a,b show our results for the sum 
of charged pions and kaons after integration over $p_T>3$ GeV, where all 
other parameters were chosen exactly as for Figs.~3a,b (since the sensitivity 
of the results to the polarized photon structure is qualitatively similar to 
the one-jet case we only consider the `maximally' saturated photon scenario 
here). One can see that the cross sections and
their asymmetries behave similarly in shape as the corresponding results
in Figs.~3a,b, but are somewhat smaller in magnitude. Nevertheless, the
expected statistical errors, calculated for the rather conservative choices 
$P_e * P_p=0.5$, ${\cal L}=100$/pb and $\epsilon=0.8$ in Eq.~(\ref{aerr}) 
and displayed in Fig.~6b, demonstrate that single-inclusive hadron 
photoproduction remains a promising candidate.
\begin{figure}[h]
\begin{center}
\vspace*{-1.2cm}
\hspace*{0.5cm}
\epsfig{file=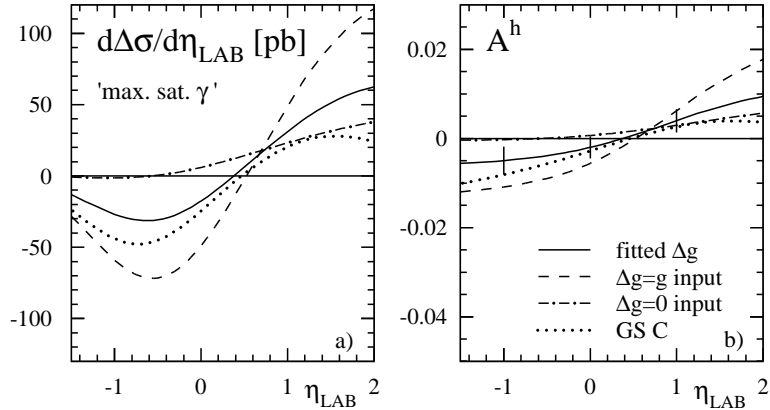,width=11.5cm}
\vspace*{-0.7cm}
\caption{\sf {\bf a,b:} Same as Figs.~3a,b, but for the case of 
single-inclusive charged hadron production, integrated over $p_T>3$ GeV.}
\end{center}
\vspace*{-1.1cm}
\end{figure}
\section{NLO Corrections to Polarized Single-Inclusive Hadron
Photoproduction}
\noindent
One major uncertainty concerning our LO results presented in the previous
section is expected to reside in the NLO corrections and the extent by
which they affect the cross sections and spin asymmetries relevant for 
experimental measurements. Only when the corrections are reasonably small 
and under control can a process that shows good sensitivity to, say, 
$\Delta g$ at the lowest order, be regarded as a genuine probe of the 
polarized gluon distribution and be reliably used to extract it from future 
data. The first basic ingredient for an extension of our results to NLO 
has been provided in the past two years by the fact mentioned above 
that NLO fits to polarized DIS data have been performed, yielding 
spin-dependent nucleon parton distributions evolved to NLO accuracy. 
Focusing on the direct part of single-inclusive hadron photoproduction, 
the calculation of the polarized cross section to NLO is then completed by 
using also (unpolarized) NLO fragmentation functions for the produced 
hadron (as provided in \cite{bkk}), and by including the ${\cal O} 
(\alpha_s)$ corrections to the spin-dependent direct subprocess cross
sections 
for the inclusive production of a certain parton that fragments into
the hadron. The latter corrections have been obtained very recently~\cite{dv}.
Technically, they involve calculation of the virtual corrections to the 
Born graphs $\vec{\gamma} \vec{q}\rightarrow gq$, $\vec{\gamma}g\rightarrow 
q\bar{q}$ and of the $2\rightarrow 3$ contributions $\vec{\gamma} \vec{a}
\rightarrow bcd$, $a,b,c,d$ being arbitrary partons and the arrows
denoting longitudinal polarization. 

We emphasize that the direct part of its own is no longer a 
well-defined quantity beyond LO since it depends on the factorization
scheme adopted. This fact is well-known from the unpolarized case, 
in which the corrections to the direct \cite{kr} {\em and} to the resolved 
\cite{guil} contributions have all been calculated. Therefore our results 
reported here will only be the first step in a full calculation of NLO 
effects to polarized single-inclusive hadron photoproduction. Despite the 
fact that they are not complete in this sense, we believe our results
to be very important both phenomenologically and theoretically: As mentioned 
earlier, the direct component dominates for HERA kinematics at $\eta_{LAB}
\leq -0.5$. Our NLO corrections for the direct part should also already be
sufficient to shed some light on the general question of perturbative 
stability of the process considered.

Fig.~7 displays the $K$-factors for the direct part of the polarized cross 
section for single-inclusive hadron photoproduction, where
\begin{equation}
K \equiv d\Delta \sigma^{NLO}/d\Delta \sigma^{LO} \; .
\end{equation}
The LO direct cross section has been calculated as in the previous figures.
For the (scheme-dependent) NLO one we have chosen the $\overline{\mbox{MS}}$ 
scheme and used also NLO ($\overline{\mbox{MS}}$) fragmentation functions 
\cite{bkk} and spin-dependent parton distributions \cite{grsv}, as well
as the two-loop expression for $\alpha_s$. As can be seen from Fig.~7,
the $K$-factors are close to unity, implying that the NLO corrections are 
rather mild. For comparison, we also show the $K$-factor for the unpolarized 
cross section. Since both the direct \cite{kr} and the resolved 
\cite{guil} contributions can be consistently calculated to NLO in this
case, we are able to plot the full physical (scheme-independent) result.
\begin{figure}[h]
\begin{center}
\vspace*{-1.6cm}
\hspace*{0.5cm}
\epsfig{file=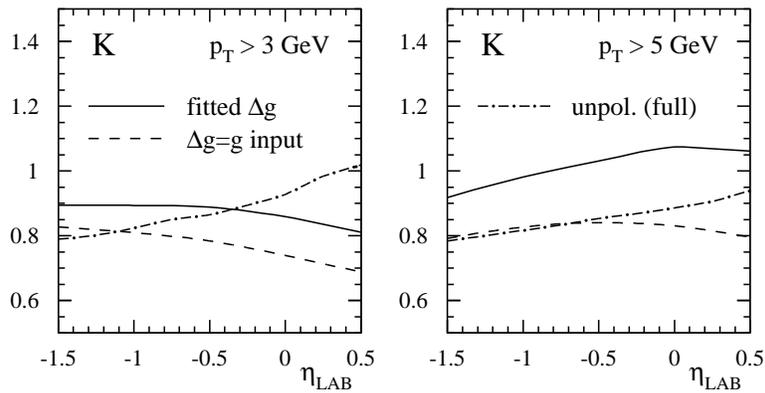,width=11.5cm}
\vspace*{-0.7cm}
\caption{\sf $K$-factors for the direct part of the polarized single-inclusive
hadron photoproduction cross section. Also shown is the $K$-factor for
the full (`direct $+$ resolved') unpolarized cross section.}
\end{center}
\vspace*{-1.1cm}
\end{figure}
 
\section{Conclusions}
\noindent
We have analyzed various photoproduction experiments in the 
context of a polarized $ep$-collider mode of HERA. We have found very
encouraging results for jet and single-inclusive hadron production which 
look promising tools for a determination of the polarized gluon distribution 
of the proton and, possibly, might even allow access to the completely 
unknown parton content of a polarized photon. We have also presented for
the first time the NLO corrections to the direct part of the polarized 
single-inclusive hadron photoproduction cross section, which reveal 
good perturbative stability. The proposed measurements 
will not be easy to do, but they seem a very interesting challenge for the 
future at HERA.

\noindent
{\bf Acknowledgements:} I am grateful to D.\ de\ Florian and M.\ Stratmann
for a fruitful and pleasant collaboration.

\end{document}